\documentclass[acus]{jac2000} 
\usepackage{epsfig}
\def\gappeq{\mathrel{ \rlap{\raise.5ex\hbox{$>$}}
                      {\lower.5ex\hbox{$\sim$}}  } }
\def\lappeq{\mathrel{ \rlap{\raise.5ex\hbox{$<$}}
                      {\lower.5ex\hbox{$\sim$}}  } }
\begin{document}
\title{Simulation Studies of the NLC with Improved Ground Motion Models
\thanks{Work supported by the U.S. Department of Energy, 
Contact Number DE-AC03-76SF00515.}}
\author{A.~Seryi, L.~Hendrickson, P.~Raimondi, 
T.~Raubenheimer, P.~Tenenbaum \\
{\it Stanford Linear Accelerator Center, Stanford University, 
Stanford, California 94309 USA}
}
\maketitle
\begin{abstract}
The performance of various systems of the Next Linear Collider (NLC)
have been studied in terms of 
ground motion using 
recently developed models. In particular, the performance of 
the beam delivery system is 
discussed. Plans to evaluate the operation of the main linac 
beam-based alignment and feedback systems are also outlined.
\end{abstract}

\section{Introduction}

Ground motion is a limiting factor in the performance of 
future linear colliders because it continuously misaligns the 
focusing and accelerating elements.
An adequate mathematical model of ground motion 
would allow prediction and optimization of the performance of 
various subsystems of the linear collider. 

The ground motion model 
presented in \cite{slacmodel} is based on 
measurements performed at the SLAC site and incorporates
fast wave-like motion, and diffusive and systematic slow 
motion. The studies presented in this paper include, 
in addition, several representative
conditions with different cultural noise contributions. 
These models were then used 
in simulations of the NLC final focus and the 
main linac. 

\section{Ground motion models}

The ground motion model for the SLAC site \cite{slacmodel} 
is based on measurements of fast motion taken
at night in one of the quietest locations in the SLAC, 
sector 10 of the linac \cite{ZDR}.

\begin{figure}[h]
\vspace{-0.6cm}
\hspace{-1.2cm}
\centering
{\vbox{
\epsfig{file=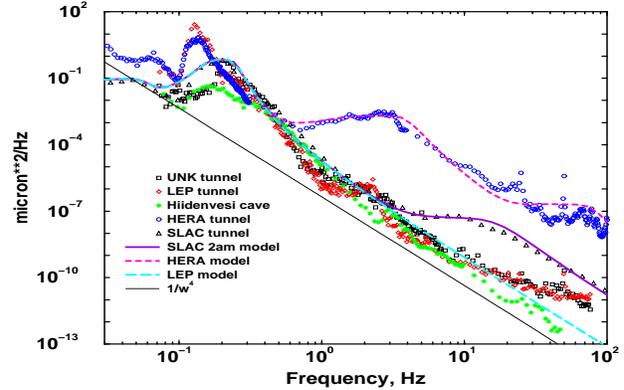,height=1.09\columnwidth,width=0.72\columnwidth,angle=-90}
}}
\vspace{-.22cm}
\caption{Power spectra measured in several places in 
different conditions \cite{vjlep,vshera,ZDR,seft} and the 
approximation curves.}
\vspace{-.5cm}
\label{meas1}
\end{figure}

To evaluate different  
levels of cultural noise, we augment this model to 
represent two other cases with significantly higher and 
lower contributions of cultural noise. 
The corresponding measured spectra and the approximations
used in the models are shown in Fig.\ref{meas1}. 

The ``HERA model'' is based on measurements in DESY \cite{vshera}
and corresponds to a very noisy shallow tunnel located in a highly 
populated area where no precautions were made to 
reduce the contribution of various noise sources in the lab
and in the tunnel. The ``LEP model'' corresponds to a deep tunnel 
where the noise level is very close to the natural seismic level, 
without additional cultural sources outside or inside 
of the tunnel. The ``SLAC model'' represents a shallow tunnel
located in a moderately populated area with a dead zone 
around the tunnel to allow damping of cultural noise and 
with some effort towards proper engineering of the in-tunnel equipment. 
(Note: the names of these models were used 
for convenience, and not to indicate the acceptability
of each particular location.)

The correlation properties of the ``LEP model'' correspond 
to a phase velosity $v=3000$~m/s \cite{vjlep}. Both the ``SLAC model'' 
and the ``HERA model'' use a phase velosity corresponding 
to $v(f) = 450+1900\exp(-f/2)$ 
(with $v$ in m/s, $f$ in Hz) which was 
determined approximately in the SLAC correlation measurements \cite{ZDR}.
This approximation was found to be suitable for 
representing the DESY correlation measurements 
\cite{vshera}, at least for frequencies greater than 
a few Hz, which contain most of the effects of the cultural noise.

\section{Applications to FFS}

The ground motion models developed were applied to 
two versions of the NLC Final Focus, 
to the one described in Ref.\ \cite{ZDR} as well as 
the current FFS described in Ref.\ \cite{newffs}. 
The FF performance is usually evaluated 
using the 2-D spectrum $P(\omega,k)$ given
by the ground motion model plus spectral response 
functions which show the contribution to the beam distortion 
at the IP of different spatial 
harmonics of misalignment. 

\begin{figure}[t]
\vspace{.03cm}
\centering
{\vbox{
\epsfig{file=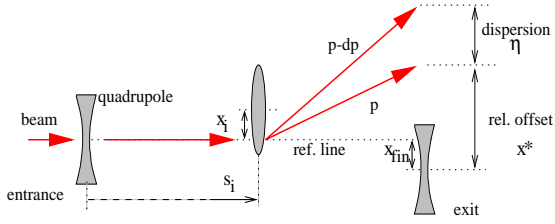,width=0.90\columnwidth,height=0.37\columnwidth}
}}
\vspace{-.04cm}
\caption{Schematic showing how quad misalignments result in the 
beam offset and dispersion.}
\vspace{.206cm}
\label{scheme}
\end{figure}

\begin{figure}[h]
\vspace{-.4cm}
\centering
{\vbox{
\epsfig{file=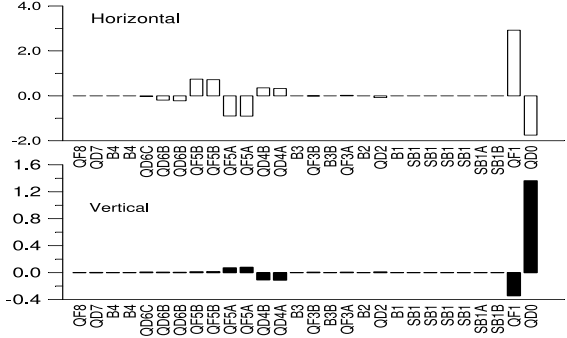,height=0.90\columnwidth,width=0.54\columnwidth,angle=-90}
}}
\vspace{-.3cm}
\caption{Coefficients $c_i = dx_{\mathrm{IP}}/dx_i$ for the new 
NLC Final Focus.
Computed using FFADA program \cite{ffada}.}
\vspace{-.4cm}
\label{coef}
\end{figure}

We summarize below the basics of the approach developed in 
\cite{seft,sn} and \cite{ZDR}. Considering a beamline
with misaligned elements, as in Fig.\ref{scheme}, 
the beam offset at the exit of the beamline and the dispersion (for example) 
can be evaluated using 
\\[-4.0mm]
$$
   x^*(t) =  \sum_{i=1}^{N} c_i \, x_i(t) - x_{\mathrm{fin}}
\,\,\, \mbox{and} \,\,\,\,\,\,
   \eta(t) =  \sum_{i=1}^{N} d_i \, x_i(t)
$$
\\[-4mm]
where $c_i =dx^*/d x_i$ and $d_i=d\eta /d x_i$ 
are the coefficients  
found using the parameters of the focusing elements and the 
optical properties of the channel. In a thin lens 
approximation to linear order, $c_i = - K_i \, r_{12}^i$ and
$d_i =  K_i \, ( r_{12}^i - t_{126}^i)$.
Here $K_i$ is $r_{21}$ of the quad matrix,
and $r_{12}^i$ and $t_{126}^i$ are the matrix elements
from the $i$-th quadrupole to the exit.
Fig.\ref{coef} shows the $c_i$ coefficients calculated 
for the new NLC Final Focus \cite{newffs}.


\begin{figure}[t]
\vspace{0.13cm}
\centering
{\vbox{
\epsfig{file=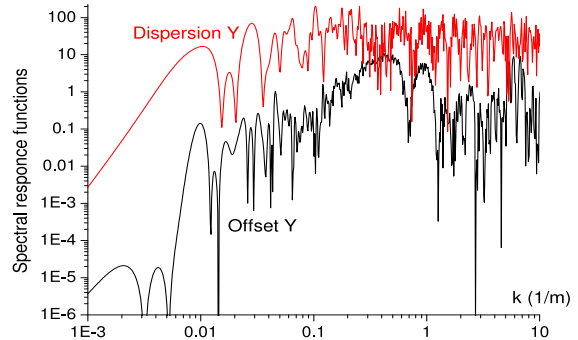,height=0.90\columnwidth,width=0.54\columnwidth,angle=90}
}}
\vspace{-.4cm}
\caption{Spectral responce functions of New NLC FF.}
\vspace{-.12cm}
\label{speacfun}
\end{figure}

It is straightforward then to combine these coefficients 
into the spectral response functions which show the 
contribution of misalignment spatial harmonics 
to the relative beam offset or to the beam distortion at the IP. 
For example, for the dispersion:
\\[-3.5mm]
$$
G_{\eta}(k) =
\left(\! \sum \limits_{i=1}^{N} d_i ( \cos(k s_i) -1) \! \right)^2 \!\! +
\left(\! \sum \limits_{i=1}^{N}  d_i \sin(k s_i) \!\right)^2
$$
\\[-3.5mm]
The spectral functions for the relative beam offset, longitudinal 
beam waist shift or coupling can be found in a similar manner 
and examples of the spectral functions for the new NLC FF are shown
in Fig.\ref{speacfun}. 

The time evolution of the beam dispersion, 
without the effect of feedbacks, can then be evaluated using 
\\[-2.5mm]
$$
\langle \eta^2 (t) \rangle =
 \int_{- \infty}^{\infty}
 P(t , k) \, G_{\eta}(k) \, \frac{dk}{2\pi}
$$
\\[-3.5mm]
where $P(t,k)$ represents a $(t,k)$ incarnation of the 
ground motion spectrum $P(\omega , k)$:
\\[-2.5mm]
$$
P(t , k) =
 \int_{- \infty}^{\infty}
P(\omega , k) \,
\,\, 2\, [1 - \cos (\omega t) ] \, \frac{d\omega}{2\pi}
$$
\\[-3.5mm]
In the case where a feedback with a gain of $F(\omega)$ is applied, 
the equilibrium beam offset can be evaluated as
\\[-2.5mm]
$$
\langle {\Delta x^*}^2 \rangle \approx
 \int_{- \infty}^{\infty}
 \int_{- \infty}^{\infty}
P(\omega , k) \,
F(\omega) \, G(k)
\, \frac{d\omega}{2\pi}
\, \frac{dk}{2\pi}
$$
\\[-3.5mm]
though more realistic simulations would be necessary 
to produce a reliable result. 
In the examples given below, we used 
an idealized approximation of the feedback gain 
function $F(\omega)= \mathrm{min}((f/f_0)^2, 1)$ with $f_0=6$~Hz;
this is a good representation of the
SLC feedback algorithm for 120~Hz operation.

\begin{figure}[h]
\vspace{-.13cm}
\centering
{\vbox{
\epsfig{file=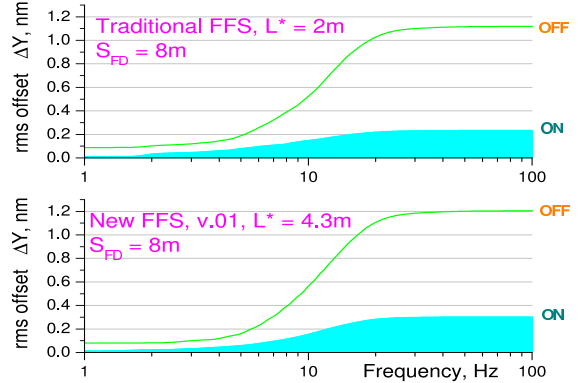,width=0.90\columnwidth,height=0.63\columnwidth}
}}
\vspace{-.4cm}
\caption{Integrated spectral contribution to the rms equilibrium 
IP beam 
offset for the traditional and new Final Focus for the 
SLAC 2AM ground motion model. Idealized rigid 
supports of the final doublets are assumed to be connected to 
the ground at $\pm S_{\mathrm{FD}}$ from the IP. 
The relative motion of the final doublets is completely 
eliminated in the case ``ON''. Red arrow shows the region 
of frequency giving the largest contribution to the rms offset.}
\vspace{-.52cm}
\label{eval1}
\end{figure}

Such analytical evaluation of ground motion, using the 
$P(\omega,k)$ spectrum and the spectral response functions 
for the transport lines is included in the PWK module of 
the final focus design and analysis code FFADA \cite{ffada}.

Evaluation of the traditional and new Final Focus in terms of 
the rms beam offset for the ``SLAC model'' is shown in Fig.\ref{eval1}. 
One can see that in terms of generalized tolerances these 
two systems are very similar. However, in the new system 
which has longer $L^*$, more rigid support can be used for the final doublet
which makes the performance closer
to the ideal. One can also see that if one could
eliminate the contribution from the final doublet by active 
stabilization, it would remove about 80\% of the effect. 

\begin{figure}[t]
\vspace{.13cm}
\centering
{\vbox{
\epsfig{file=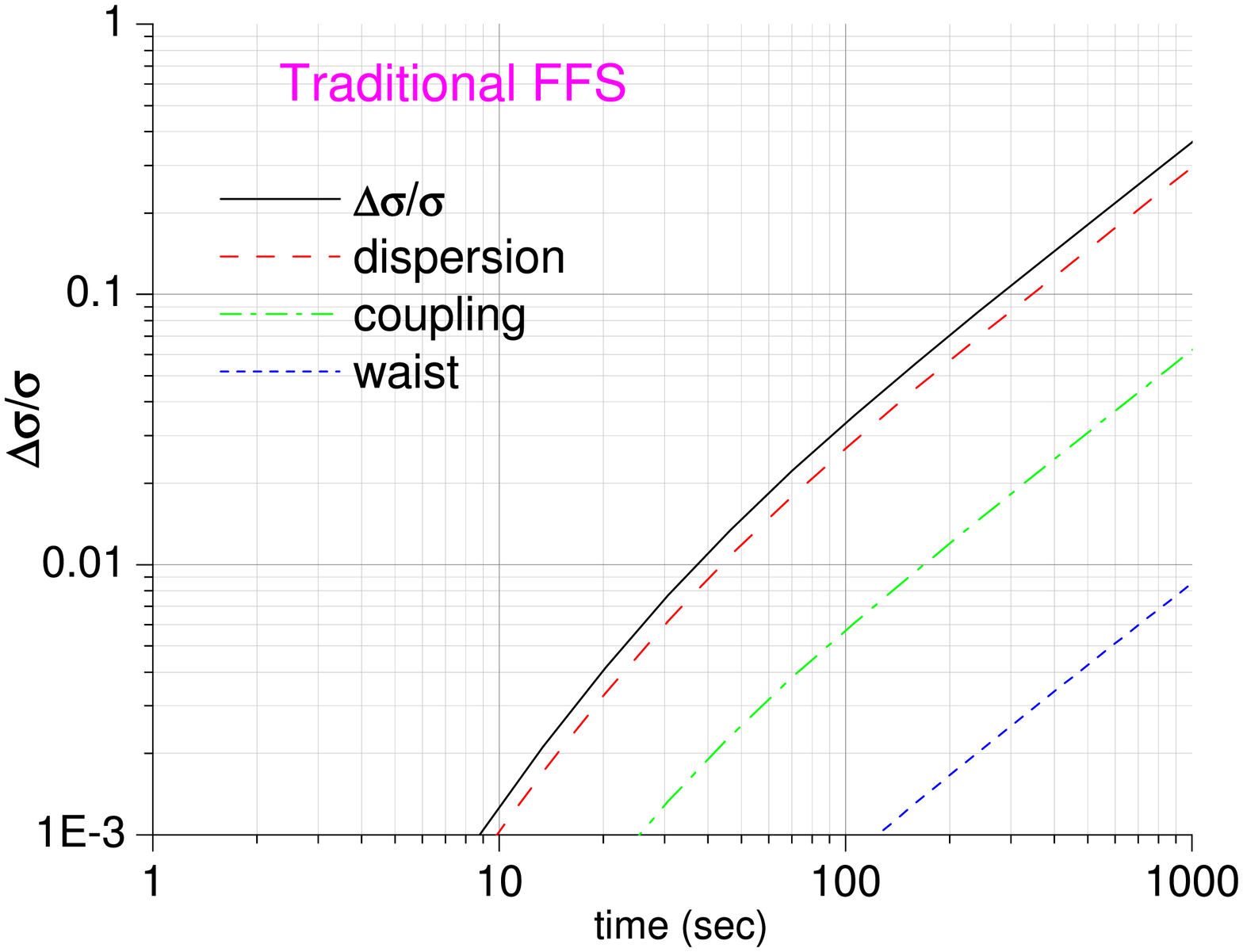,height=0.90\columnwidth,width=0.52\columnwidth,angle=90}
\epsfig{file=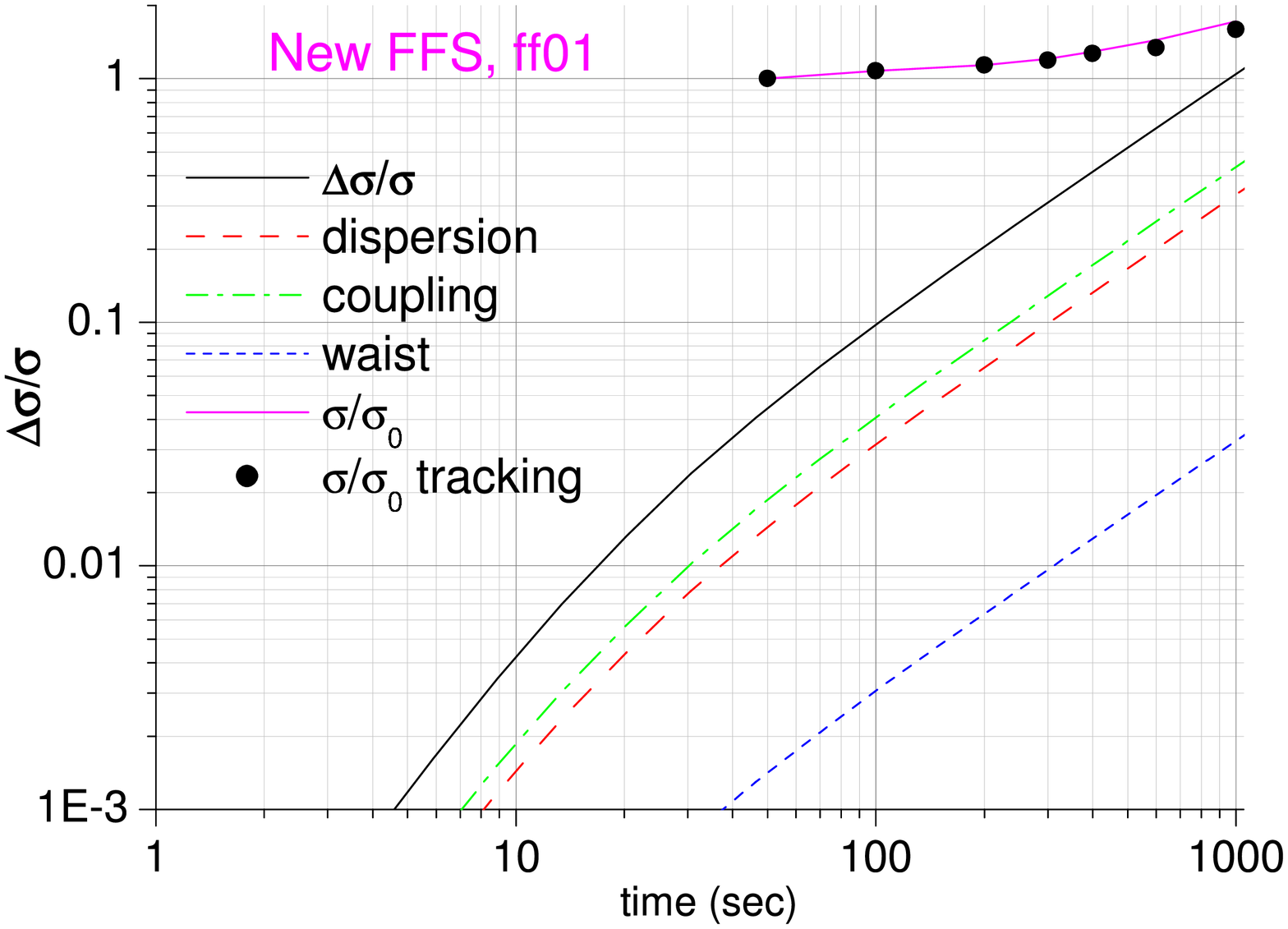,height=0.90\columnwidth,width=0.52\columnwidth,angle=90}
}}
\vspace{-.4cm}
\caption{Beam distortion at the IP for the traditional and new NLC FF
versus time for the ``SLAC model'' of ground motion, free evolution. 
Note that orbit feedback would drastically decrease this beam distortion. 
Results were computed using the FFADA program \cite{ffada}. }
\vspace{-.6cm}
\label{distort}
\end{figure}

The free IP beam distortion evolution for the traditional and 
new NLC FF is shown in Fig.\ref{distort} for the ``SLAC model''.  
Note that an orbit correction which could keep the orbit stable 
through the sextupoles would drastically decrease 
this beam distortion. The picture presented is therefore
useful only for comparison of the performance of the two FF systems. 
One can see, that the new FF, having longer $L^*$ and correspondingly 
higher chromaticity, has somewhat tighter tolerances. 
The orbit feedback, however, may be much simpler since there are fewer 
sensitive elements in the new system.

The analytical results presented in Fig.\ref{distort} are in 
good agreement with the tracking. One should note here that 
the tracking was done with an energy 
spread which is 3 times smaller than nominal (see \cite{newffs} for 
these beam parameters) because otherwise the second
order tracking routine of the MONCHOU program used for 
misalignment simulation did not 
produce reliable results when compared with other programs.

\begin{figure}[b]
\vspace{-.45cm}
\centering
{\vbox{
\epsfig{file=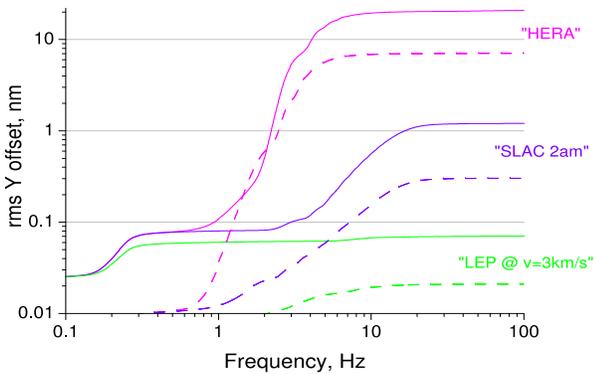,height=0.95\columnwidth,width=0.59\columnwidth,angle=90}
}}
\vspace{-.4cm}
\caption{Integrated spectral contribution to the rms equilibrium 
IP beam offset for the new Final Focus with FD supports at 
$S_{\mathrm{FD}} =\pm 8$~m for different models 
of ground motion. Dashed curves correspond to the
complete elimination of relative motion of the final 
quads. }
\vspace{-.02cm}
\label{models}
\end{figure}

Comparison of the performance of the new FF in terms of 
different ground motion models is shown in Fig.\ref{models}. 
One can see that a site located in a highly populated area
without proper vibration sensitive engineering would present 
significant difficulties for a linear collider with the 
parameters considered. Stabilization of only the final doublet would not 
be sufficient in this case. A site with noise similar 
to the ``SLAC model'' would certainly be suitable, 
while the ``LEP model'' would be suitable even for much more
ambitious beam parameters. These results should not be considered 
as an attempt to evaluate any particular site, or even the models, 
because for a fully consistent assessment, various in-tunnel 
noise sources as well as 
vibration compensation methods must be considered together. 

\begin{figure}[t]
\vspace{.13cm}
\centering
{\vbox{
\epsfig{file=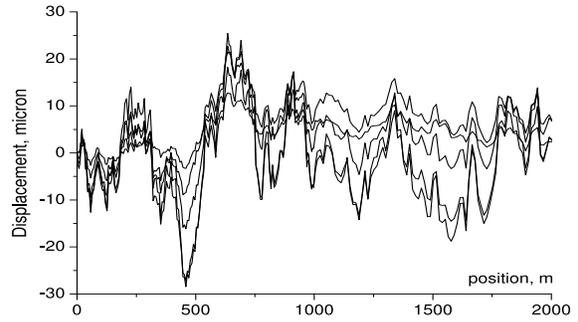,height=0.90\columnwidth,width=0.50\columnwidth,angle=90}
}}
\vspace{-.4cm}
\caption{LIAR generated misalignments of a linac
for ``SLAC model'' and $\Delta T=8$~hours between curves. }
\vspace{-.5cm}
\label{liar1}
\end{figure}

\section{Applications to linac}

The models now developed, which more adequately 
describe the various components of ground motion, can also be 
applied to simulations of the beam based alignment 
procedures and cascaded feedback in the main linac. 
Such simulations require direct modeling of 
misalignments which is done by summing 
harmonics whose amplitudes are given by 
the 2-D spectrum of the corresponding ground motion model.
In this case, since a large range of $T$ and $L$
must be covered in a single simulation run,
the harmonics are distributed over the relevant 
$(\omega,k)$ range equidistantly in a 
logarithmic sense \cite{sl96}. Such a method of ground 
motion modeling is now included in the 
linear accelerator research code LIAR \cite{liar} 
in addition to the previously implemented ATL model.
An example of the misalignments generated by LIAR 
is shown in Fig.\ref{liar1}.

\section{Conclusion}

New ground motion models 
now incorporate various sources of ground motion 
such as wave-like motion, diffusive and systematic motion. 
These models are being used to evaluate and optimize 
performance of various subsystems of the NLC.

\end{document}